\documentclass[useAMS,usenatbib,referee]{mn2e}

\newcommand{\apj}{ApJ}
\newcommand{\apjl}{ApJL}

\newcommand{\mnras}{MNRAS}

\newcommand{\aap}{A\&A}

\newcommand{\apjs}{ApJS}

\usepackage{epsfig}

\usepackage{graphicx}

\title[Seismology of ZZ Cetis]{Seismological studies of ZZ Ceti stars. I. The
model grid and the application to individual stars.}

\author[B. G. Castanheira and S. O. Kepler]
{B. G. Castanheira${^{1,2}}$ and S. O.
Kepler${^1}$ \\
$^1$Departamento de Astronomia, Universidade Federal do Rio Grande do Sul, 
Av. Bento
Gon\c{c}alves 9500\\ Porto Alegre 91501-970, RS, Brazil\\
$^2$ Instit\"ut f\"ur Astronomie, T\"urkenschanzstr. 17, A-1180 Wien, Austria}

\begin{document}

\pagerange{\pageref{firstpage}--\pageref{lastpage}} 

\maketitle

\label{firstpage}

\begin{abstract}
We calculate and explore an extensive adiabatic model grid for pulsating
white dwarfs with H dominated atmospheres, the ZZ Ceti stars. We also compared
the computed modes with the observed ones for five ZZ Ceti stars that are
a representative sample of the whole class of pulsators. We describe our new
approach for seismological studies, using the relative observed amplitudes to
give weights for the periods in the fit and the external mass and temperature
determinations as a guide. Our seismological study is clear evidence that
seismology is indeed a powerful tool in the study of stellar structure and
evolution.
\end{abstract}


\section{Introduction}
\label{Intro}

Because of the $\kappa$-$\gamma$ mechanism and/or convection driving, white
dwarfs pulsate in three different instability strips, depending on the chemical
element that drives pulsation. The 142 currently known DAVs (or ZZ Ceti stars)
comprise the most numerous class of pulsating white dwarfs. These stars have
hydrogen atmospheres and are confined in an observational narrow instability 
strip between 10\,800 and 12\,300\,K. It is essential to note that pulsation is
a phase every white dwarf goes through, i.e., the properties we measure for 
pulsating stars apply to white dwarfs as a whole. The other two classes are the
DBVs, with helium-rich atmospheres, and the DOVs, the hotter pulsators, with
roughly a dozen of known pulsators in each class. Their temperatures range from
20\,000 to 30\,000\,K and 75\,000 to 170\,000\,K, respectively. 

Pulsating white dwarfs constitute the largest population of variable stars, even
though we can only detect the nearest ones because of their intrinsic faintness.
Despite the relatively small number of known pulsators, their studies have
extremely important astrophysical implications. Pulsations probe the internal
structure of white dwarfs, as each pulsation mode yields an independent 
measurement of their interiors. Fundamental properties of the stars, such as
mass, can be derived from just a few pulsation periods (Bradley \& Winget 1994,
Fontaine et al. 1991). If many modes are excited, the structure and composition
of the white dwarfs can be determined in detail and with high precision (Kepler
et al. 2003, Winget et al. 1994). Since they constitute the evolutionary end 
point of 95--98\% of all stars, information on their internal structure and
composition provide a critical test for stellar evolution models, including the
double shell burning phase at the AGB and mass loss rates. As the periods of a
pulsating white dwarf change slowly as the star cools, their rate of change
measure the stellar cooling rate (e.g. Kepler et al. 2005), allowing the 
calibration of theoretical calculations. The cooling rates constrain in turn
such properties as the rate of crystallization in the white dwarf core (Kanaan
et al. 2005, Winget et al. 1997), plasmon neutrino interaction cross sections 
(Kawaler et al. 1986, Winget et al. 2004), and even axion masses (C\'orsico et
al. 2001; Kim, Montgomery, \& Winget 2006). Coupled with the white dwarf 
luminosity function, the cooling rates yield a star formation history for the
galactic disk and, in principle, the age of the oldest disk and halo components
(Winget et al. 1987, Hansen et al. 2002, 2007, Kalirai et al. 2007). 

The ZZ Ceti stars can be separated in three groups: hot, intermediate, and cool.
The hot ZZ Ceti stars have only a few modes excited to visible amplitudes, all 
of them with short periods (smaller than 300\,s), small amplitudes (from 1.5 to
20\,mma), and the power spectrum does not change substantially from season to
season. These stars define the blue edge of the ZZ Ceti instability strip. The
cool ZZ Ceti stars are at the opposite side of the strip, showing a very rich
pulsational spectra of high amplitude modes (up to 30\%) and longer periods (up
to 1500\,s). The modes interfere with each other, causing dramatic changes in
the pulsational spectra, but in general, the total amount of energy transported 
by pulsation is conserved. There are also a few red edge ZZ Ceti stars with long
periods of low amplitude, representing the stage when the star is ceasing to 
pulsate. The intermediate ZZ Ceti stars have characteristics between the hot and
the cool groups.

In this paper, we start by a detailed investigation on the seismological model
grid, followed by the application of a new method of seismology to five ZZ Ceti
stars. These stars are a representative sample of the whole class of the ZZ Ceti
stars.

\section{The seismological models}

\subsection{Calculating the models}

The starting point to calculate the evolutionary model is a quasi-static 
pre-white dwarf model. The internal structure of white dwarfs with 
$T_{\mathrm{eff}}\sim12\,000\,$K has been sufficiently modified due to cooling,
stellar contraction, and chemical diffusion, so that using a hot polytrope 
static model is equivalent to using models evolved from all the previous phases.
The cooling sequence of those polytropes converge to the same structure obtained
from self-consistent models even at temperatures above the observed pulsational
instabilities for DAs (e.g. Wood 1990).

We used the White Dwarf Evolutionary Code described in details by Lamb \& van 
Horn (1975) and Wood (1990), to evolve the starting model until the desired
temperature. This code was originally written by Martin Schwarzschild, but has
been modified and updated by many astronomers (Montgomery 1998 and references 
therein). The
equation of state for the core of our models is from Lamb (1974) and for the
envelopes from Fontaine, Graboske, \& van Horn (1977). We used the updated 
opacity OPAL tables (Iglesias \& Rogers 1996), neutrino rates of Itoh et al.
(1996), and ML2/$\alpha$=0.6 mixing length theory of Bohm \& Cassinelli (1971).
The core evolution calculations are self-consistent, but the envelope was 
treated separately. The core and the envelope are stitched together, 
fulfilling the
boundary conditions in the interface. The transitions between the layers are
consistent with time diffusion, following Althaus et al. 
(2003), especially for the transition zone between hydrogen (H) and helium (He).

\subsection{Grid dimensions}

We calculated an extensive adiabatic model grid for the pulsation modes, varying
four quantities: $T_{\mathrm{eff}}$, $M$, $M_{\mathrm{H}}$, and 
$M_{\mathrm{He}}$. The whole grid was generated from a starting polytrope of
$M=0.6\,M_{\odot}$, using homology transformations to obtain models with masses
from 0.5 to 1.0\,$M_{\odot}$, in steps of 0.005\,$M_{\odot}$, initially. The
effective temperature ($T_{\mathrm{eff}}$) varies from 10\,600 to 12\,600\,K, in
steps of 50\,K. The upper H and He layer mass values are $10^{-4}\,M_*$ and
$10^{-2}\,M_*$, respectively; above these limits, nuclear reactions would excite
g-mode pulsations in the PNNVs by the $\epsilon$ mechanism, which are not 
observed (Hine 1988). 
The lower H and He layer mass values are $10^{-9.5}\,M_*$ (the minimum
H amount to be detected in a DA white dwarf spectra) and $10^{-3.5}\,M_*$,
respectively.

\subsection{Core composition: C/O}

\begin{figure}
\resizebox{\hsize}{!}{\includegraphics[angle=0]{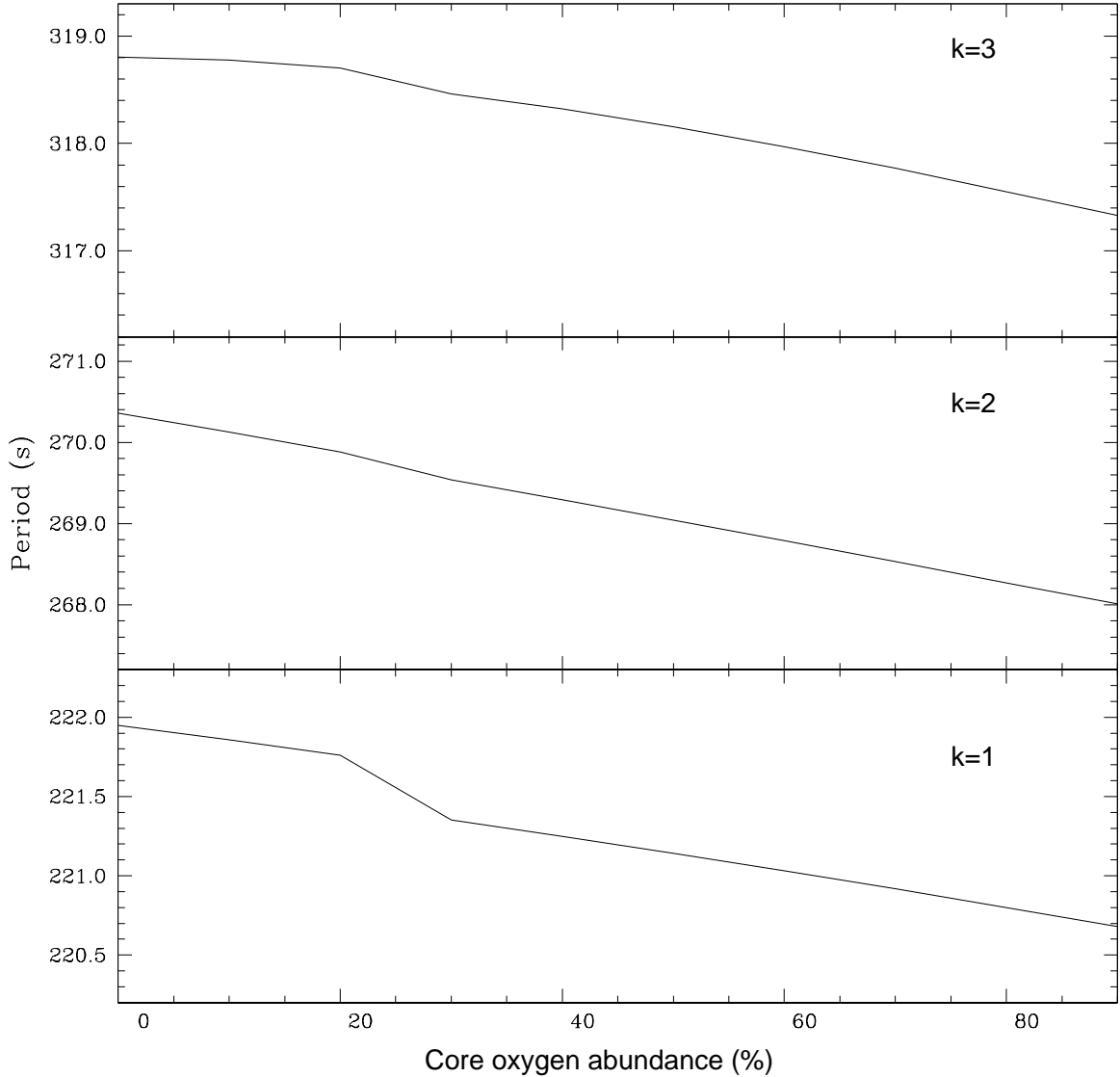}}
\caption{Variation of the three first $\ell=1$ overtones changing the core
abundances, for a model with $T_{\mathrm{eff}}=11\,600$\,K, $M=0.6\,M_{\odot}$,
$M_{\mathrm{H}}=10^{-4}\,M_*$, and $M_{\mathrm{He}}=10^{-2}\,M_*$. In the lower
panel is the variation for the $k=1$ mode, in the middle panel for $k=2$, and 
in the upper panel for $k=3$.} 
\label{graf_co}
\end{figure}

Varying the core composition introduces three parameters in the fit: the
abundance itself from the uncertain C($\alpha$,$\gamma$)O reaction rate 
(Metcalfe 2005) and the two extreme points of the function that better describes
the transition zone. We chose a homogeneous core of 50:50 C/O, for 
simplicity. 
In figure~\ref{graf_co}, we show the variation in period with O abundance for
the three first $\ell=1$ overtones. We calculated a representative model of a
star in the middle of the instability strip, $T_{\mathrm{eff}}$=11\,600\,K, 
with mass close to the DA white dwarfs mean distribution, $M=0.6M_{\odot}$ (e.g.
Kepler et al. 2007), and canonic values for H and He layer masses,
$M_{\mathrm{H}}=10^{-4}M_*$ and $M_{\mathrm{He}}=10^{-2}M_*$. The change is only
a few seconds when the O core abundance varies from 0\% to 90\% (the 
complementary quantity is C)!

\begin{figure}
\resizebox{\hsize}{!}{\includegraphics[angle=0]{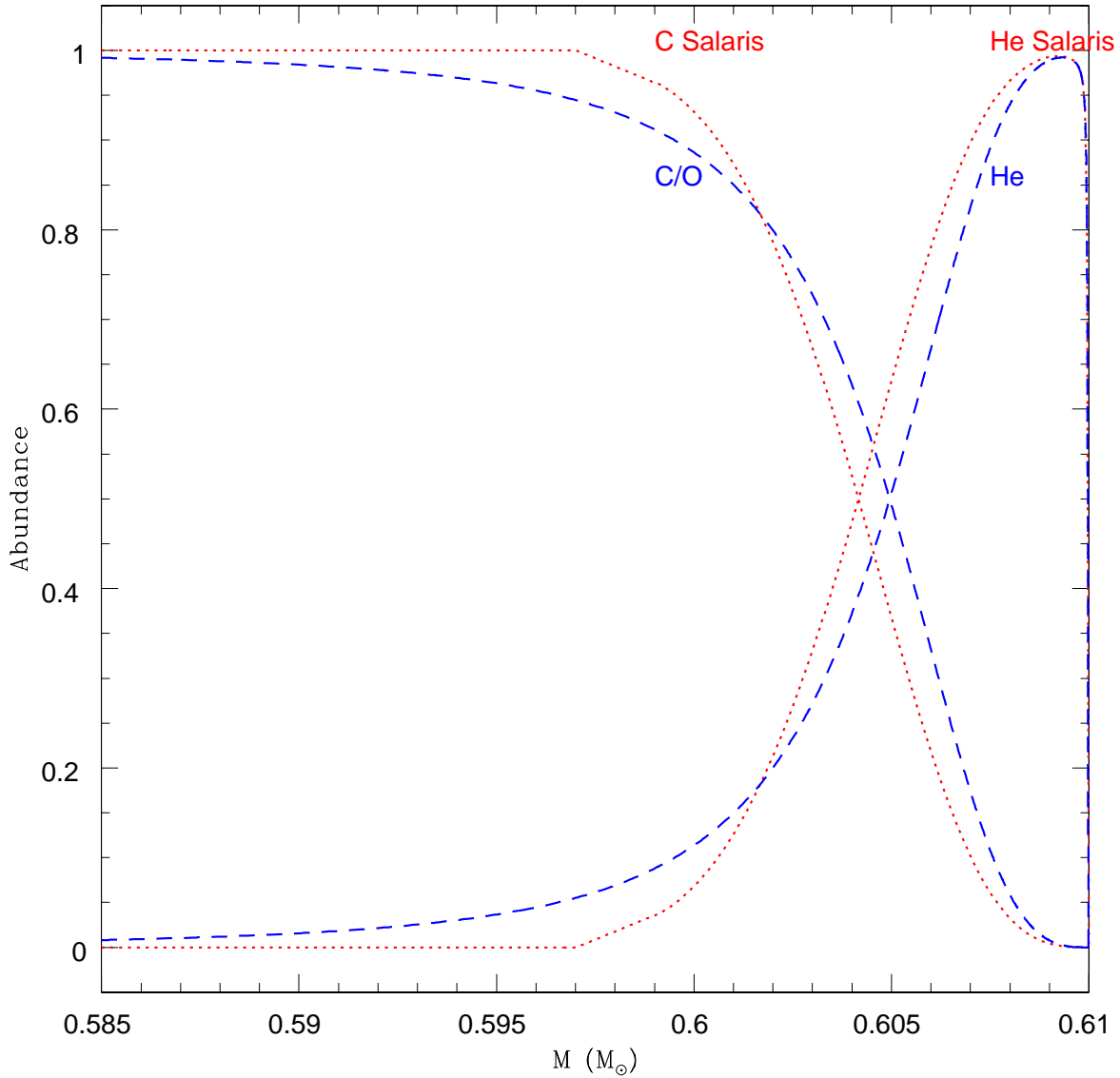}}
\caption{Comparison between Salaris profile (dotted red line) and our 
choice of homogeneous C/O=50:50 core (dashed blue line). The position of the
transitions are different, causing differences in the calculated modes, which 
are trapped within these zones. However, the differences can be compensate by 
changing the thickness of the He layer.}
\label{graf_core}
\end{figure}

We also investigated the shape of the transition zones when a different profile
is used. In figure~\ref{graf_core}, we show the abundance variation as function
of mass. The Salaris profile (Salaris et al. 1997) was derived from 
self-consistent models, and the simple profile is the one we used in our main
grid. Both models were calculated for $T_{\mathrm{eff}}=12\,000\,$K, 
$M=0.61\,M_{\odot}$, $M_{\mathrm{H}}=10^{-4}\,M_*$, and 
and $M_{\mathrm{He}}=10^{-2}\,M_*$. Clearly, the transition zones between the
core and the He outer envelope are in different positions; therefore, the 
trapped modes in these cavities are necessarily different. Because the ZZ Cetis
pulsate with only a few modes, we decided to use a fixed homogeneous C/O 50:50
core to decrease the number of model parameters, but still be consistent with
the reaction rate uncertainty. The price payed for this choice is that the He
layer mass determinations are uncertain. The differences in the shape of the
transition zone can be compensated by changes in the thickness of the He layer.
The Salaris profile introduces a more complicated than a linear decrease
of the O profile in the outer layers (see Fig. 2. of Metcalfe 2005). This
effect potentially changes the calculated trapped modes in the models. However,
because the ZZ Cetis pulsate in only a few modes, we did not include the
changes in the core profile in our fit.

For high and low mass white dwarfs, core composition plays an important role and
should be slightly different, but these stars are a small fraction ($\leq30$\%)
of all known ZZ Cetis, and white dwarfs in general. In the few cases that mass
determinations are precise enough, it is necessary to compute models for other 
C/O proportions (see section 3.5).

\subsection{Exploring the model grid}

\begin{figure}
\resizebox{\hsize}{!}{\includegraphics[angle=0]{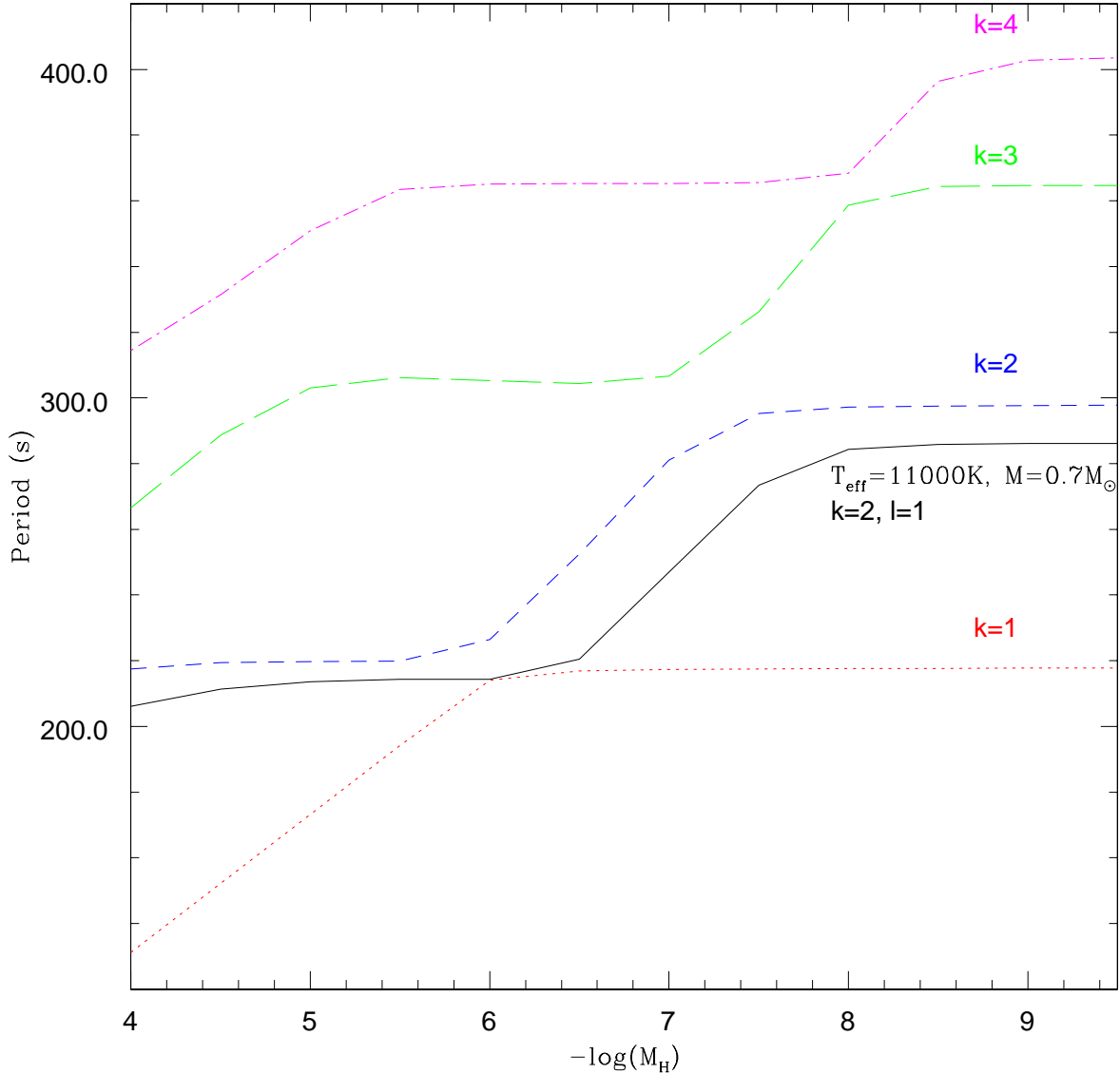}}
\caption{Variation of the first $\ell=1$ overtones with the mass of the H 
envelope for a model with $T_{\mathrm{eff}}=12\,000$\,K, $M=0.6\,M_{\odot}$, 
and $M_{\mathrm{He}}=10^{-2}\,M_*$. The dotted (red) line is for the mode $k=1$,
the dashed (blue) line for $k=2$, the long-dashed (green) line for $k=3$, and 
the dotted-dashed (magenta) line for $k=4$. The continuous (black) line is the
$k=2$ $\ell=1$ mode variation, but for a model with 
$T_{\mathrm{eff}}=11\,000$\,K, $M=0.7\,M_{\odot}$, and 
$M_{\mathrm{He}}=10^{-2}\,M_*$. Both models are for a homogeneous C/O=50:50
core.}
\label{fig2}
\end{figure}

The ZZ Ceti stars at the blue edge of the instability strip pulsate with only a 
few short period modes, the first overtones. We show the changes in the period 
of the first overtone as function of the H layer mass thickness in 
figure~\ref{fig2}, for a model with $T_{\mathrm{eff}}=12\,000$\,K, 
$M=0.6\,M_{\odot}$, and $M_{\mathrm{He}}=10^{-2}\,M_*$.

The first observation is the presence of avoided crossings in the models, which
occurs when a mode assumes the properties of the immediate next $k$ mode. The
consequence is a variable period spacing ($\Delta P$), depending on the 
thickness of the H layer.

Consider a hypothetical star that pulsates with an $\ell=1$ mode at 217.5\,s.
For the model with $T_{\mathrm{eff}}=12\,000$\,K, $M=0.6\,M_{\odot}$, and
$M_{\mathrm{He}}=10^{-2}\,M_*$, there are two possible solutions:
$M_{\mathrm{H}}=10^{-7.5}\,M_*$ if $k=1$ or $M_{\mathrm{H}}=10^{-4}\,M_*$ if
$k=2$ (see figure~\ref{fig2}). On the other hand, if the atmosphere parameters 
have a different value, for instance, $T_{\mathrm{eff}}=11\,000$\,K and
$M=0.7\,M_{\odot}$ (solid line in figure~\ref{fig2}), the best solution would be
$M_{\mathrm{H}}=10^{-6.5}\,M_*$, if the mode is $\ell=1$ $k=2$. Nevertheless, a
$T_{\mathrm{eff}}=11\,000$\,K solution can be excluded, because this is a
typical temperature of a red edge pulsator, while the observed period is typical
of a blue edge pulsator. This simple example illustrates the degeneracy of our
model grid. It is thus necessary to detect as many modes as possible to find an
unique solution for the stellar structure of our targets.

\begin{figure}
\resizebox{\hsize}{!}{\includegraphics[angle=0]{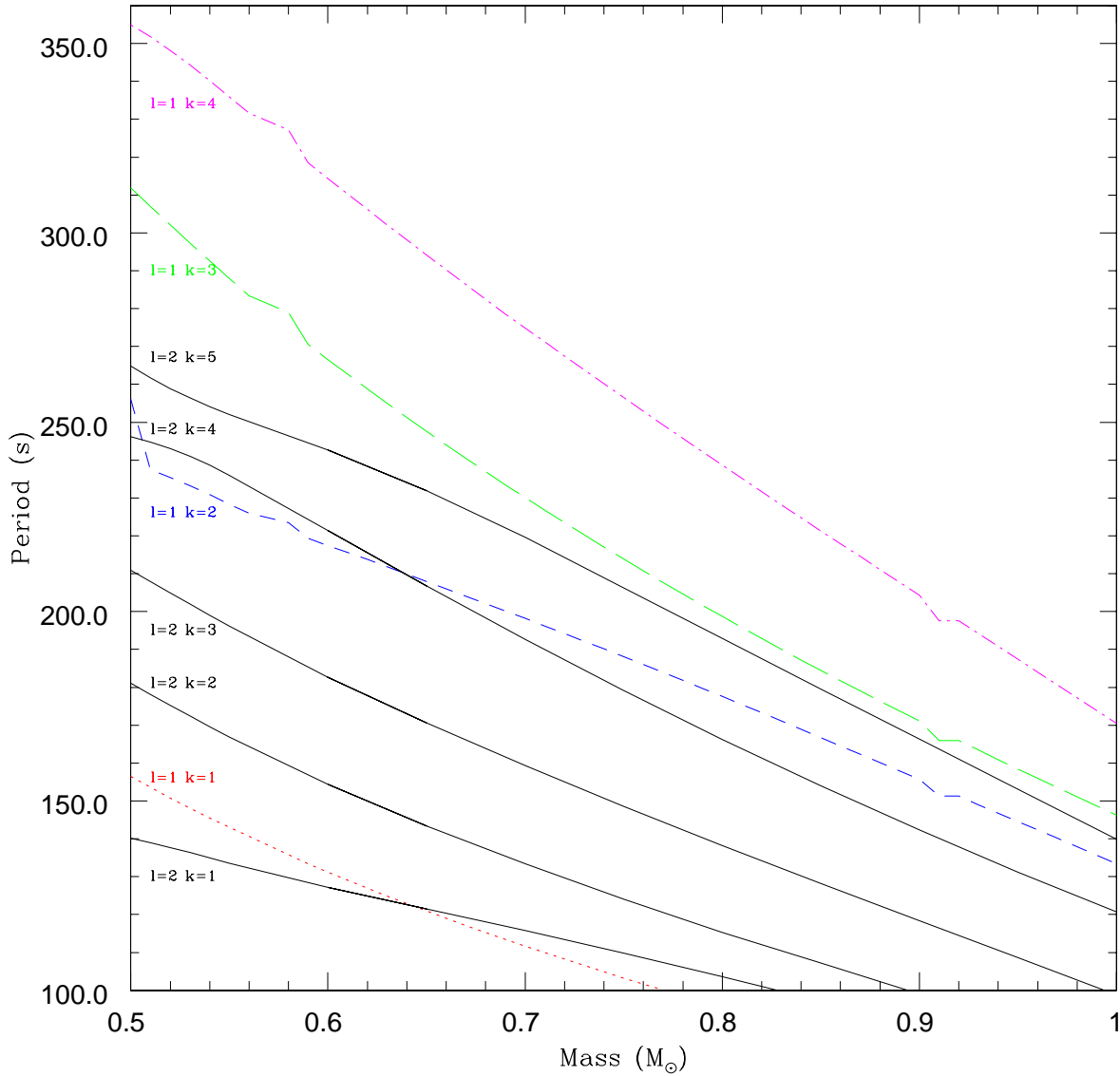}}
\caption{Variation of the first overtones for a model with 
$T_{\mathrm{eff}}=12\,000$\,K, $M_{\mathrm{H}}=10^{-4}\,M_*$, and
$M_{\mathrm{He}}=10^{-2}\,M_*$ as a function of mass. The solid lines are 
for models with $\ell=2$ and the other lines are for $\ell=1$ modes. The
first $k$s are for shorter periods and the higher, for longer periods. 
For the same $\ell$ and $k$ mode, the modes are shorter for higher mass
values.}
\label{fig3}
\end{figure}

In figure~\ref{fig3}, we show the changes of the first $\ell=1$ and 2 overtones 
with stellar mass, for a model with $T_{\mathrm{eff}}=12\,000$\,K,
$M_{\mathrm{H}}=10^{-4}\,M_*$, and $M_{\mathrm{He}}=10^{-2}\,M_*$. For the
previous $\ell=1$ 217.5\,s pulsator, the possible solutions are:
0.60\,$M_{\odot}$ if $k=2$, 0.74\,$M_{\odot}$ if $k=3$, and 0.86\,$M_{\odot}$
if $k=4$.

We also computed models for $\ell=2$, 3, and 4 modes, despite the fact they are 
less likely to be observed in the optical, because of the geometrical
cancellation (Robinson, Kepler, \& Nather 1982). Robinson et al. (1995), Kepler 
et al. (2000), and Castanheira et al. (2004) studied the chromatic amplitude
variation and conclude that most of the observed ZZ Cetis modes are $\ell=1$. 
The results
of Kotak, van Kerkwijk, \& Clemens (2004) and Clemens, van Kerkwijk, \& Wu 
(2000) point to the same conclusion, even though their Keck optical data sampled
only a small wavelength range. That is an important point, because there are
more excitable $\ell=2$ modes than $\ell=1$ in any model (see 
figure~\ref{fig3}). If our hypothetical star pulsates with an $\ell=2$ mode 
instead, the best fits would be: 0.61\,$M_{\odot}$ if $k=4$ and 
0.71\,$M_{\odot}$ if $k=5$.

\begin{figure}
\resizebox{\hsize}{!}{\includegraphics[angle=0]{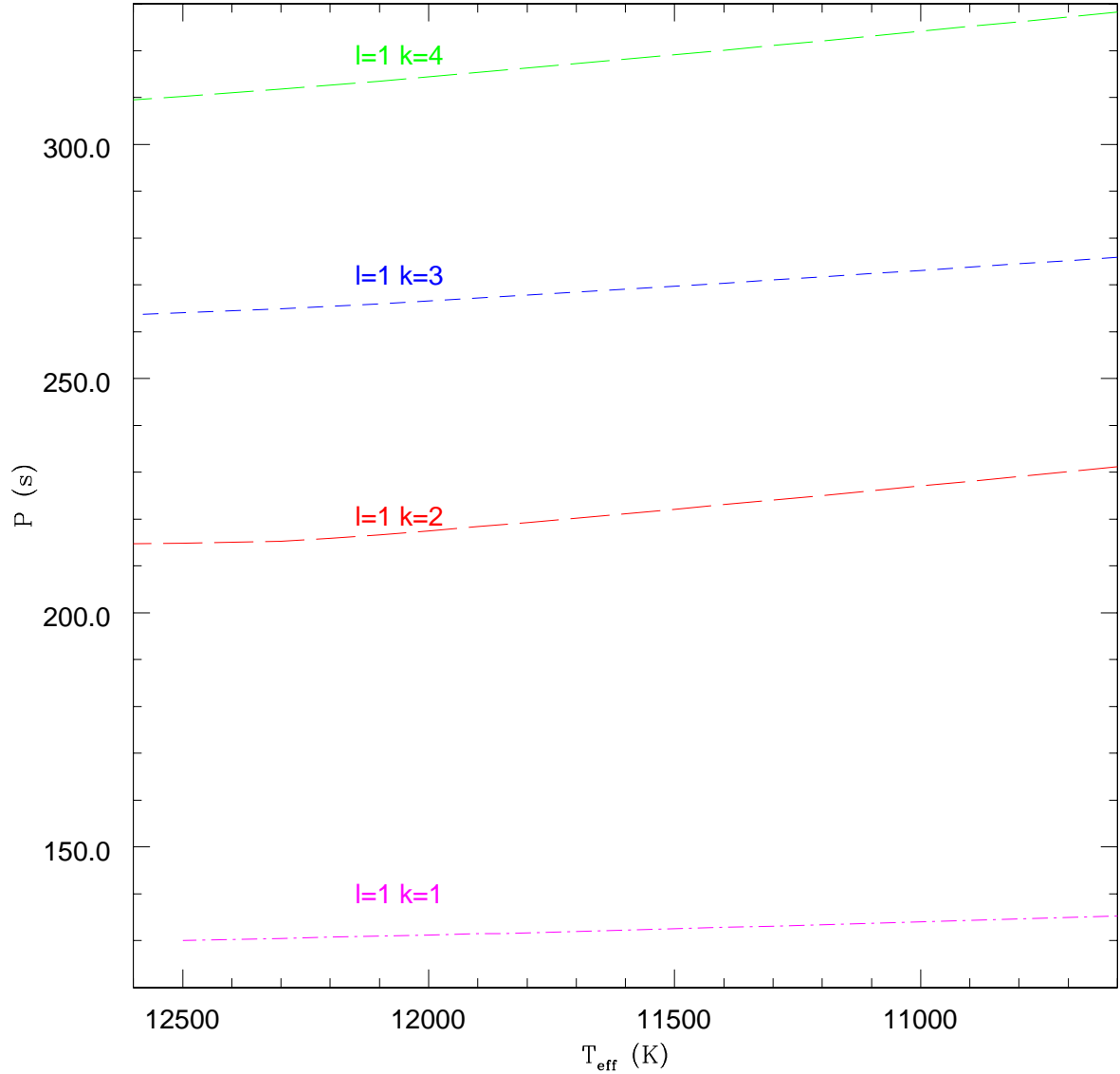}}
\caption{Variation of the first overtones for a model with $M=0.6\,M_{\odot}$,
$M_{\mathrm{H}}=10^{-4}\,M_*$, and $M_{\mathrm{He}}=10^{-2}\,M_*$, as a 
function of $T_{\mathrm{eff}}$.}
\label{fig4}
\end{figure}

Figure~\ref{fig4} shows the variation of the first $\ell=1$ overtones as a
function of $T_{\mathrm{eff}}$, for models with $M=0.6\,M_{\odot}$, 
$M_{\mathrm{H}}=10^{-4}\,M_*$, and $M_{\mathrm{He}}=10^{-2}\,M_*$. The most
important result is that the modes are getting longer as the models cool down.
Physically, we are measuring the core cooling, which is the dominant effect in
the ZZ Ceti phase. 

\begin{figure}
\resizebox{\hsize}{!}{\includegraphics[angle=0]{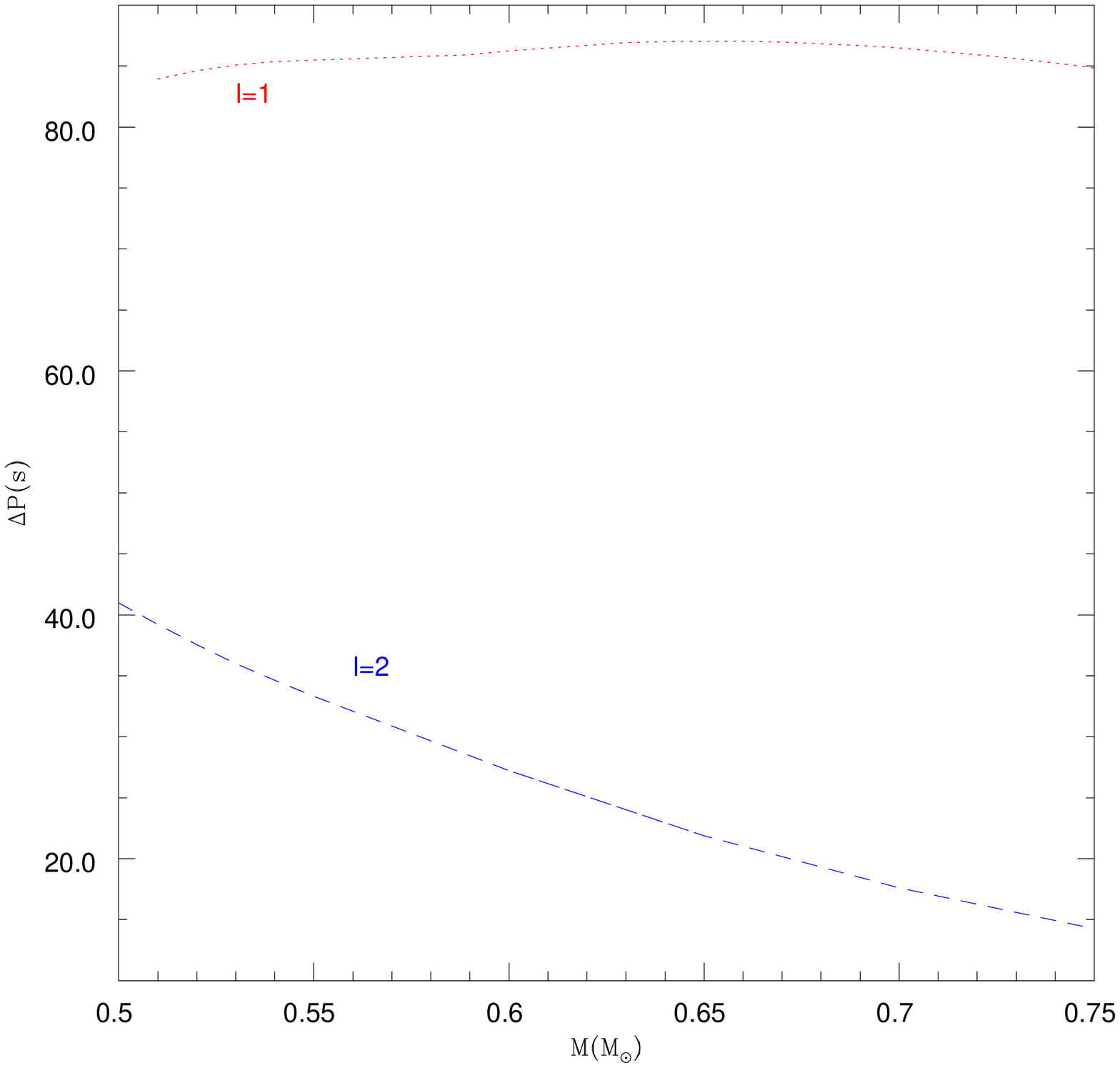}}
\caption{Period spacing between the two first overtones ($k=1$ and 2) for 
$\ell=1$ and 2 modes. The difference between higher $\ell$ modes is smaller
in comparison to smaller $\ell$ modes, for a particular model.}
\label{fig8}
\end{figure}

At the blue edge, the lower $k$s are excited (Winget, van Horn, \& Hansen 1981
and Brassard et al. 1991). In figure~\ref{fig8}, we show the period spacing
$\Delta P$ between the two first overtones ($k=1$ and 2) for $\ell=1$ and 2.
For stars with $M\sim0.6\,M_{\odot}$, if close modes could not be explained by
avoided crossings, a period spacing smaller than around 50\,s should indicate
$\ell=2$.

\section{Seismology of individual stars}

\subsection{G117-B15A: our test star}

Since the discovery of its variability (McGraw \& Robinson 1976), G117-B15A
was considered the most regular ZZ Ceti. Its light curve is nearly sinusoidal,
showing mean light variations with $\simeq$ 2\% amplitude. Due to the mode
stability, Kepler et al. (2005) measured the rate of change of the main mode at
215\,s, with a precision better than 10$^{-15}$\,s/s. G117-B15A is the most
precise optical clock known.

G117-B15A pulsates with only three independent low amplitude modes, showing also
the harmonic of the main mode and linear combinations between the modes (Kepler
et al. 1982). In table~\ref{tab1}, as an example, we list the average values 
for all 
periodicities detected in observations with the Argos camera (Nather \& Mukadam
2004) at the 2.1\,m telescope at McDonald Observatory.

\begin{table}
\begin{tabular}{ccc}\hline\hline
Period (s) & Amplitude (mma) & identification \\ \hline
215.20 & 17.36 & f1 \\
270.46 & 6.14 & f2 \\
304.05 & 7.48 & f3 \\ \hline
107.7 & 1.65 & 2$\times$f1 \\
126.2 & 1.40 & f1+f3 \\
119.8 & 1.30 & f1+f2 \\ \hline
\end{tabular}
\caption{All detected periodicities for G117-B15A, with mode labeled as
well as the linear combinations and the harmonic of the main mode, which are
not fundamental modes of the star.}
\label{tab1}
\end{table}

We used G117-B15A as a test of our seismological analysis for two main reasons.
First, this star shows a very simple pulsation spectrum, but more than one mode
is excited. Second, there are published seismological studies, which can be used
to test the validity of our approach.

Our seismological study starts with a literature search for all available 
information about the star. Besides more than 30 years of time series photometry
(Kepler et al. 2005), there are independent measurements of the atmospheric
properties ($T_{\mathrm{eff}}$ and $\log g$) of G117-B15A, from different 
techniques, summarized in table~\ref{tab2}. The differences among the published
values can be explained by the intrinsic degeneracy between $T_{\mathrm{eff}}$
and $\log g$, which allows more than one combination of solutions.

\begin{table}
\begin{tabular}{ccccc}\hline\hline
Method & $T_{\mathrm{eff}}$ (K) & $\log g$ (cgs units) & $M$ ($M_{\odot}$) & 
Spectroscopic ref. \\ \hline
Optical spectra LPT & 11\,630$\pm$200 & 7.97$\pm$0.05 & 0.59$\pm$0.03 &
Bergeron et al. 2004 \\
UV spectra + V mag. & 11\,900$\pm$140 & 7.86$\pm$0.14 & 0.53$\pm$0.07 &
Koester \& Allard 2000 \\
Gravitational reddening & & 8.00$\pm$0.14 & 0.58$\pm$0.08 & Wegner \& Reid 
1991 \\ \hline
\end{tabular}
\caption[]{Independent atmospheric determinations for G117-B15A.}
\label{tab2}
\end{table}

As our models do not account for the excitation of linear combinations nor
harmonics, it is very important to be sure that only normal modes are being
used in the fit. For G117-B15A, the modes are identified in table~\ref{tab1} as
$f_1$, $f_2$, and $f_3$. To perform seismology, we compared the observed modes
($P_{\mathrm{obs}}$) with the computed ones ($P_{\mathrm{model}}$) by adding the
square of the differences, similar to a $\chi^2$ fit, according to the 
expression: 

\begin{equation}
S=\sum_{i=1}^n \sqrt{\frac{[P_{\mathrm{obs}}(i)-P_{\mathrm{model}}]^2\times 
w_P(i)}{\sum_{i=1}^n w_P(i)}}
\end{equation}
where $n$ is the number of observed modes and $w_P$ is the weight given to 
each mode.

In order to give higher weight to the highest amplitude mode, we choose to
weight the periods by the energy they transport, which is proportional to the
observed amplitude square ($A^2$). Even though our model grid uses the adiabatic
approximation and does not provide theoretical amplitudes, we are using the
observed relative amplitudes to weight our fits, to guarantee that the fit 
will be always dominated by high amplitude modes. For G117-B15A, the periods and
their respective normalized weights are 215.20\,s, 270.46\,s, and 
304.05\,s, and 0.13, 1.00, and 0.67.

At first, we have tried $\ell=1$ for all modes. This choice is supported by
previous observations of Robinson et al. (1995) and Kotak, van Kerkwijk, \&
Clemens (2004). 
We would have tried higher $\ell$ values if no solution was found, despite the
fact that all chromatic amplitude changes to date indicate the modes are more
likely $\ell=1$ than higher $\ell$ for all ZZ Ceti stars (e.g. Kepler et al. 
2000).

\begin{figure}
\resizebox{\hsize}{!}{\includegraphics[angle=0]{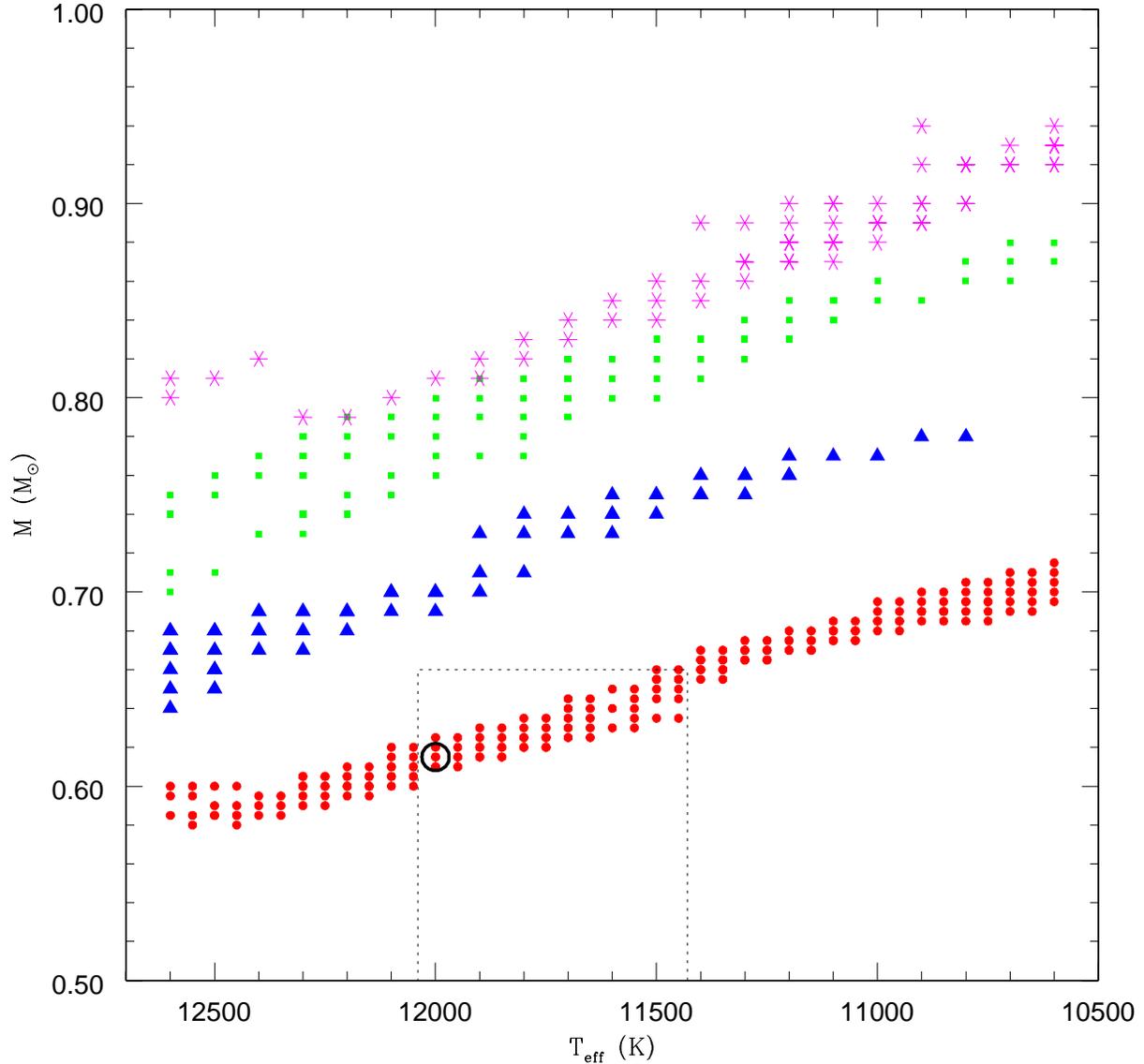}}
\caption{Results from comparison between the pulsation modes of the star
G117-B15A and the models. The (red) circles are the solutions for 
$M_{\mathrm{He}}=10^{-2}M_*$, the (blue) triangles for 
$M_{\mathrm{He}}=10^{-2.5}M_*$, the (green) squares for
$M_{\mathrm{He}}=10^{-3}M_*$, and the (magenta) asterisks for
$M_{\mathrm{He}}=10^{-3.5}M_*$. The dotted line box limits the region of the
independent temperature and mass determinations ($\pm 1\sigma$) 
listed in table~\ref{tab2} and
the open circle shows the location of the minimum for this family of solutions.}
\label{fig9}
\end{figure}

By comparing the observations and the models, we found several families of
solution below the cut $S<1.8$\,s (the quadratic sum of the relative 
uncertainties). Figure~\ref{fig9} shows the possible combinations of
$T_{\mathrm{eff}}$, $M$, $M_{\mathrm{H}}$, and $M_{\mathrm{He}}$ below the
$S$ cut off, and the minimum of each family is listed in table~\ref{tab3}.

\begin{table}
\begin{tabular}{||c|c|c|c|c|c|c||}\hline\hline
Symbol in fig.~\ref{fig9} & $T_{\mathrm{eff}}$ (K) & $M$ ($M_{\odot}$) &
$-\log M_{\mathrm{H}}$ & $-\log M_{\mathrm{He}}$ & $S$ (s) & Modes ($\ell,k$)\\
\hline\hline
(Red) Circles & {\bf 12\,000} & {\bf 0.615} & {\bf 7} & {\bf 2} & 0.67 &
215.3(1,1), 275.0(1,2)\\
&&&&&& 302.1(1,3)\\ \hline
(Blue) Triangles & 11\,500 & 0.75 & 5 & 2.5 & 0.97 & 215.1(1,2), 265.1(1,3)\\
&&&&&& 308.0(1,4)\\ \hline
(Green) Squares & 12\,600 & 0.71 & 7.5 & 3 & 0.73 & 215.4(1,1), 266.0(1,2) \\
&&&&&& 301.6(1,3)\\ \hline
(Pink) Asterisks & 11\,500 & 0.85 & 8.5 & 3.5 & 0.19 & 215.2(1,1), 271.5(1,2)\\ 
&&&&&& 303.3(1,4) \\ \hline
\end{tabular}
\caption[]{Seismological solutions for G117-B15A: absolute minima for each
possible family of solution in figure~\ref{fig9}.}
\label{tab3}
\end{table}

Zhang, Robinson, \& Nather (1986) derived the following expression to calculate
the uncertainties in the fit parameters:

\begin{equation}
\sigma^2=\frac{d^2}{S-S_0}
\end{equation}
where $d$ is the imposed step, $S_0$ is the absolute minimum, and $S$
is the local minimum in a $d$ difference between the quantities.

The mean uncertainties for the quantities we fit are 
$\sigma_{T_{\mathrm{eff}}}\sim50\,$K, $\sigma_M\sim0.005\,M_{\odot}$,
$\sigma_{M_{\mathrm{H}}}\sim10^{-0.5}\,M_*$, and 
$\sigma_{M_{\mathrm{He}}}\sim10^{-0.5}\,M_*$, of the order of the spacings in
the model grid. These values are the typical uncertainties in our seismological
studies.

The final step is to identify which of the possible solutions correlates with
the atmospheric determinations. From table~\ref{tab2}, the mass should be 
within the
range 0.46\,$M_{\odot}$ to 0.66\,$M_{\odot}$, which excludes all seismological
high mass 
solutions; only the first solution in table~\ref{tab3} is consistent with the
previous atmospheric determinations. In the same way, the spectroscopic
temperature is within the range 11\,430 to 12\,040\,K, which also agrees with
this solution. Our seismological solution indicates that $T_{\mathrm{eff}}$ is
hotter than determined by Bergeron et al. (2004), but consistent with the
value from the UV spectra (see table~\ref{tab2}).

Alternatively to our procedure, Bradley (1998) applies his six step recipe. The
first step is to select seismological models consistent with spectroscopic mass
and temperature. He uses a standard model with $M_{\mathrm{He}}=10^{-2}\,M_*$
and C/O nominal profile. Afterwards, he adjusts $M_{\mathrm{H}}$ to obtain a 
mode with $\ell=1$, $k=1$ or 2 close to 215\,s. Then, he refines 
$T_{\mathrm{eff}}$ and $M_{\mathrm{H}}$ to bring the 270\,s mode into agreement.
The next step is to use the 304\,s mode to adjust C/O profile. Finally, he 
refines the adjustment even more and/or applies small changes to 
$M_{\mathrm{H}}$, $M_{\mathrm{He}}$, and in the core structure. He found that 
the mass of the H layer is either $M_{\mathrm{H}}=10^{-4}\,M_*$ or 
$10^{-7}\,M_*$, depending if the 215\,s $\ell=1$ mode is $k=2$ or 1,  
respectively. His best result in mass is 0.60\,$M_{\odot}$, which agrees with
spectroscopic values.

The code he used to calculate his models is basically the same as we used. 
However, his version used the trace element approximation, while our updated 
version uses a parametrization that mimics the results of
time-dependent diffusion calculations (Althaus et al. 2003)
to describe the transition zones. The
biggest advantage of our seismological recipe is that we can explore all 
possibilities, avoiding local minima and error propagation if the spectroscopic 
determinations are uncertain.

Benvenuto et al. (2002) used realistic He, C, and O profiles, calculated from
previous stellar evolutionary phases, obtaining $T_{\mathrm{eff}}=11\,800$\,K,
$M=0.525\,M_{\odot}$, and $M_{\mathrm{H}}=10^{-3.83}\,M_*$, for G117-B15A. 
Their $T_{\mathrm{eff}}$ and mass are consistent with Koester \& Allard (2000)
determinations by fitting the UV spectrum.

Our seismological study of G117-B15A is in agreement with other approaches,
encouraging us to apply the same technique to other ZZ Cetis.

\subsection{G185-32: a rebel star}

The exotic behavior of G185-32 has been noticed since the discovery of its
variability (McGraw et al. 1981). The subharmonic ($3f_0/2$) and
the second harmonic ($3f_0$) are excited and have comparable amplitudes to the
main periodicity at $P=215\,$s ($f_0$) in the optical. Castanheira et al. (2004)
identified the $\ell$ degree of at least five pulsation modes using chromatic
amplitude differences. The subharmonic at 141.9\,s was proposed to be a 
nonlinear effect and the shortest periodicity at 70.9\,s ($3f_0$) should be
$\ell=2$, probably $k=1$.

Thompson et al. (2004) studied the UV (Kepler et al. 2000) and optical chromatic
amplitude variations of the 142\,s periodicity and proposed it as an $\ell=4$ 
mode. In their model, even though $\ell=4$ mode cancels itself over the visible
hemisphere for almost all inclination values, the harmonic does not, because it
has a different surface distribution, similar to lower $\ell$ modes. In this 
scenario, the harmonic would be allowed to have a higher amplitude than the
fundamental mode.

Yeates et al. (2005) studied the amplitude ratios between harmonics and their
parent modes, for hot ZZ Ceti stars, to constrain the $\ell$ index, based on
Wu's (2001) theoretical predictions. They proposed the 141.9\,s mode as an 
$\ell=3$. 

Pech \& Vauclair (2006) suggest that G185-32 pulsates mainly with $\ell=2$ 
modes. The 72.5\,s mode was labeled by them as $\ell=2$ $k=1$ and used as 
reference mode, even though it has never been observed as the dominant mode. 
They proposed that the 70.9\,s, 148.5\,s, 181.9\,s, 212.8\,s, and 215.7\,s (the 
dominant periodicity in all data sets) are not real modes. The 141.9\,s 
periodicity would be a superposition of a real $\ell=2$ mode and the linear 
combination of two parent modes, with $\ell=3$ and $\ell=4$ or 5. The resulting
amplitude of an $\ell=2$ would be modified by the interference of the linear
combinations. In this scenario, it would not be surprising that the 142\,s 
chromatic amplitude variation does not follow the theoretical predictions. Their
analysis does not take into account the relative amplitudes and assumes that the
temperature and mass derived from optical spectra are accurate.

Bradley (2006) suggests yet another possibility: the modes at $P=72.5$\,s and
$141.9$\,s would be $\ell=4$, $k=2$ and $k=8$, respectively. However, the high
geometric cancellation of $\ell=4$ modes is not taken into account by his fits.

Clearly there is no agreement about G185-32's seismology. Therefore, we also
studied G185-32 in the same way as we did for G117-B15A, to check if our method
agrees with any other previous mode identifications. In table~\ref{tabg185}, we
show G185-32's atmospheric parameters derived from different methods.

\begin{table}
\begin{tabular}{||c|c|c|c|c||}\hline\hline
Method & $T_{\mathrm{eff}}$ (K) & $\log g$ (in cgs) & $M$ ($M_{\odot}$)
& Ref. spectroscopy \\ \hline \hline
Optical spectra & 12\,130$\pm$200 & 8.05$\pm$0.05 & 0.64$\pm$0.03 &
Bergeron et al. 2004 \\
UV spectra + V magnitude & 11\,820$\pm$200 & 7.92$\pm$0.10 & 0.57$\pm$0.05 &
Koester \& Allard 2000 \\
Many methods & 11\,960$\pm$80 & 8.02$\pm$0.04 & 0.62$\pm$0.02 &
Castanheira et al. 2004 \\ \hline
\end{tabular}
\caption{Independent atmospheric determinations for G185-32.}
\label{tabg185}
\end{table}

Our model computes only $m=0$ modes, because we did not include rotation nor
magnetic field, which would split modes with different $m$'s. Metcalfe (2003)
concluded that the families of solutions are the same even if we use the wrong
$m$ mode, for slow rotators (like pulsating white dwarfs). Once again, it is
necessary to exclude the linear combinations in the fit, which is definitely not
a trivial task for G185-32. In table~\ref{tab6}, we list the periodicities 
published by Castanheira et al. (2004), their optical amplitudes, and the linear
combinations. As the periods at 537.6\,s, 454.6\,s, and 181.9\,s were marginally
detected, we did not include them in our seismological analysis.

\begin{table}
\begin{tabular}{||c|c|c||}\hline\hline
Periodicities (s) & Amplitude (mma) & Identification \\ \hline \hline
651.70 & 0.67 & $f_1$ \\
560.77 & 0.09 & $f_3-f_1$ \\
537.59 & 0.57 & (?) \\
454.56 & 0.38 & (?) \\
370.21 & 1.62 & $f_2$ \\
300.60 & 1.04 & $f_3$\\
266.17 & 0.46 & $f_4$ \\
215.74 & 1.93 & $f_5$ \\
181.90 & 0.03 & (?) \\
148.45 & 0.57 & $f_7-f_6$ \\
141.87 & 1.43 & $f_6$ ($\sim 3f_5/2$)\\
72.54 & 0.93 & $f_7$ ($\sim 3 f_5$) \\
70.93 & 0.69 & $2\times f_6$ \\ \hline
\end{tabular}
\caption{All periodicities detected for the star G185-32, with the mode 
identification of the linear combinations and harmonics of the highest 
amplitude periodicity. The question mark was used for the modes marginally
detected, which were not used in our seismological studies.}
\label{tab6}
\end{table}

\begin{figure}
\resizebox{\hsize}{!}{\includegraphics[angle=0]{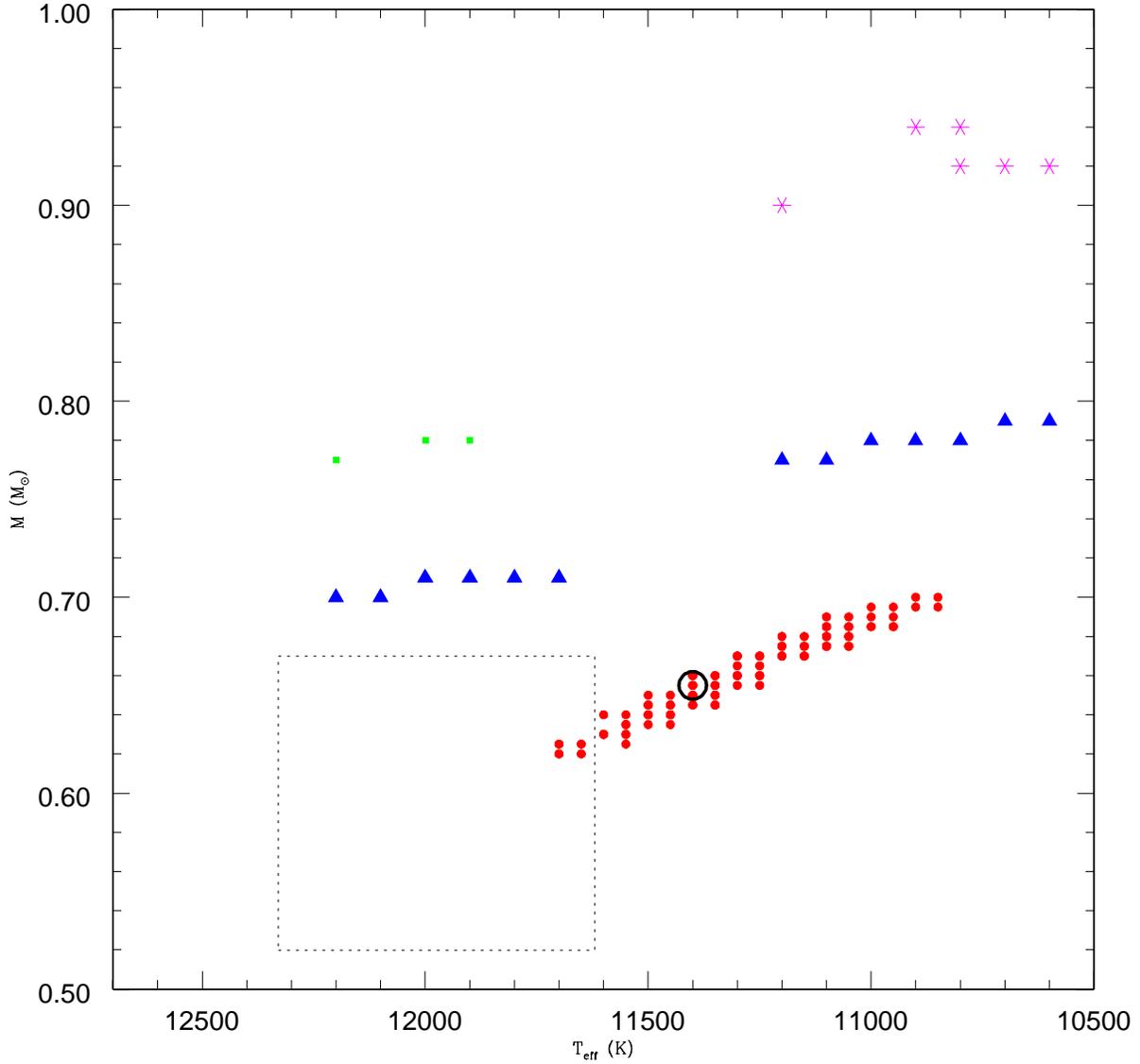}}
\caption
{Results of the fit of the pulsation modes for the star G185-32. The (red)
circles are the solutions for $M_{\mathrm{He}}=10^{-2}\,M_*$, the (blue)
triangles for $M_{\mathrm{He}}=10^{-2.5}\,M_*$, the (green) squares for 
$M_{\mathrm{He}}=10^{-3}\,M_*$, and the (magenta) asterisks for
$M_{\mathrm{He}}=10^{-3.5}\,M_*$. The dotted line box limits the region of the
independent temperature and mass determinations listed in table~\ref{tabg185}
and the open circle shows the location of the minimum for this family of 
solutions.}
\label{sol3}
\end{figure}

The first attempt was to use only the modes longer than 215\,s (from $f_1$ to
$f_5$), which is a very similar analysis to that of G117-B15A. 
In figure~\ref{sol3},
we show all the solutions for $S<2$\,s, assuming that all modes are $\ell=1$. 
The minima of the families of solutions are listed in table~\ref{tsol3}.

\begin{table}
\begin{tabular}{||l|c|c|c|c|c|c||}\hline
Symbol in the plot~\ref{sol3} & $T_{\mathrm{eff}}$ (K) & $M$ ($M_{\odot}$)
& $-\log M_{\mathrm{H}}$
& $-\log M_{\mathrm{He}}$ & $S$ (s) & Modes ($\ell,k$) \\ \hline\hline
1. Circles (red) & {\bf 11\,400} & {\bf 0,655} & {\bf 7} & {\bf 2}
& 0,75 & 216,0(1,1), 262,3(1,2)\\
&&&&&& 298,3(1,3), 369,5(1,4) \\
&&&&&& 654,5(1,10)\\ \hline
2. Triangles (blue) & 10\,900  & 0,78  & 2,5 & 5 & 0,85 & 215,1(1,2),
258,0(1,3)\\
&&&&&& 303,2(1,4), 370,4(1,6)\\
&&&&&& 649,6(1,13)\\ \hline
3. Squares (green) & 11\,900 & 0,78  & 3 & 5 & 1,94 & 216,7(1,2),
258,2(1,3)\\
&&&&&& 303,7(1,4), 370,2(1,6)\\
&&&&&& 634,9(1,12)\\ \hline
4. Asterisks (magenta) & 10\,700 & 0,92 & 3,5 & 6 & 0,93 & 215,7(1,2),
260,9(1,3)\\
&&&&&& 301,6(1,4), 368,7(1,6)\\
&&&&&& 655,0(1,14) \\ \hline
\end{tabular}
\caption{Absolute minima of the various families of solutions of the 
seismological analysis for the star G185-32.}
\label{tsol3}
\end{table}

The closest seismological solution to the previous atmospheric determinations is
solution 1: $T_{\mathrm{eff}}=11\,400$\,K, $M=0.655\,M_{\odot}$,
$M_{\mathrm{H}}=10^{-7}\,M_*$, and $M_{\mathrm{H}}=10^{-2}\,M_*$. In this model,
a 71.43\,s $\ell=4$ $k=1$ mode is present. There is also an $\ell=4$ $k=5$ 
mode at 150.9\,s, a substantial difference of 9\,s from the observed periodicity
at 141.8\,s. The lack of a mode closer to the observed value in the best model 
supports the proposal that this periodicity is, in fact, a nonlinear effect. On
the other hand, Bradley's (2006) suggestion for the 72\,s mode to be $\ell=4$
agrees with our seismological study. Despite the fact its observed amplitude is 
not consistent with an $\ell=4$ mode, because of the large geometrical 
cancellation expected, the amplitude could have been amplified by
the resonance with the harmonic of the main mode.

The best fits for G117-B15A and G185-32 indicate that these stars have a very
similar structure: $M_{\mathrm{H}}=10^{-7}\,M_*$ and 
$M_{\mathrm{He}}=10^{-2}\,M_*$. The total masses are a little different though. 
As G117-B15A has $\langle P \rangle=233.1$\,s and G185-32 has 
$\langle P \rangle=280.4$\,s, their seismological temperatures are 
consistent with the second being a little cooler than the first one (as in
Mukadam et
al. 2006).

\subsection{G226-29: poor man's seismology}
                                                                                
G226-29 is the brightest known ZZ Ceti ($V=12.22$) and the closest one too 
($d\sim12$\,pc). This star is one of the (if not the) hottest ZZ Ceti stars, 
according to temperature and mass determinations listed in table~\ref{tabg226}.

\begin{table}
\begin{tabular}{||c|c|c|c||}\hline
Method & $T_{\mathrm{eff}}$ (K) & $M$ ($M_{\odot}$) & Spectroscopic reference\\
\hline \hline
LPT optical spectra & 12\,460$\pm$200 & 0.79$\pm$0.03 & Bergeron et al. 2004\\
LPT optical spectra & 12\,260$\pm$200 & 0.81$\pm$0.03 & Gianninas et al. 2005\\
UV spectra + V mag. & 12\,050$\pm$160 & 0.73$\pm$0.07 & Koester \& Allard 2000\\
\hline
\end{tabular}
\caption{Independent determinations for the atmospheric parameters of G226-29.}
\label{tabg226}
\end{table}
                                                                                
On the other hand, G226-29 is the poorest star in terms of observed modes. 
Kepler et al. (1995) reported the detection, even on a Whole Earth Telescope 
(WET) campaign, of a single triplet, with periods very close to 109\,s. The
central component is at 109.278\,s, the mode we used to study the internal
structure of this star. The presence of a triplet is consistent with rotational
splitting of an $\ell=1$ mode. The $\ell=1$ identification is also reinforced by
the UV chromatic amplitude variation (see Kepler et al. 2000 for further 
discussions).

\begin{figure}
\resizebox{\hsize}{!}{\includegraphics[angle=0]{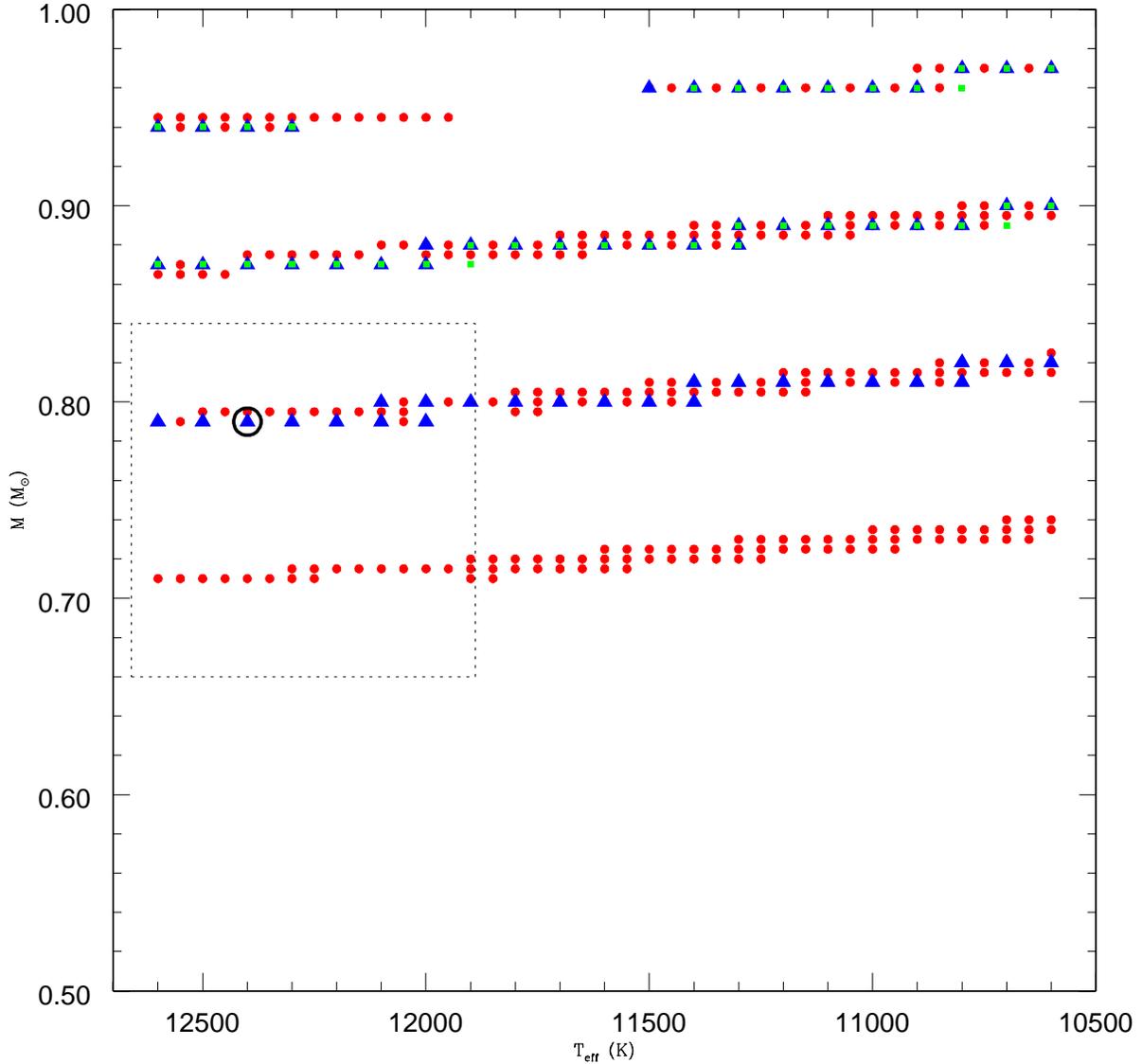}}
\caption{Fit results for the pulsation mode of the star G226-29. The (red)
circles are the solutions for $M_{\mathrm{He}}=10^{-2}\,M_*$, the (blue)
triangles for $M_{\mathrm{He}}=10^{-2.5}\,M_*$, and the (green) squares for
$M_{\mathrm{He}}=10^{-3}\,M_*$. There is no solution for 
$M_{\mathrm{He}}=10^{-3.5}\,M_*$. The conclusion is that the stellar mass should
be high and that the He layer should be thick.
The dotted line box limits the region of the
independent temperature and mass determinations listed in table~\ref{tsol2} and
the open circle shows the location of the minimum for this family of solutions.}
\label{sol2}
\end{figure}

The total uncertainty used in our fit is $S$=1\,s, because the models are not
more accurate than this, due to the uncertainties in the physical parameters,
even though the observations for this star are more precise. In 
figure~\ref{sol2}, we show the families of solutions for different He layer 
masses, identified by different symbols.

Even with only one excited mode, there are no solutions for masses lower than 
0.7\,$M_{\odot}$! This result shows that seismology is indeed a powerful tool to
study stellar interiors and that some information can be obtained even
from a single mode.

Using the previous atmospheric determinations, the mass range is between 0.66
and 0.84\,$M_{\odot}$ and temperature, 11\,890 and 12\,660\,K. In
table~\ref{tsol2}, we list the minima of the only two families of solution 
within this interval. The two possible seismological solutions differ by
0.005\,$M_{\odot}$ in mass and 300\,K in temperature. The H layer can only be
thick and around $10^{-4.5}\,M_*$, and the He layer mass can only be between
$10^{-2}\,M_*$ and $10^{-2.5}\,M_*$ (modulo the core profile).

\begin{table}
\begin{tabular}{||l|c|c|c|c|c|c||}\hline
Symbol in plot~\ref{sol2} & $T_{\mathrm{eff}}$ (K) & $M$ ($M_{\odot}$)& 
$-\log M_{\mathrm{H}}$ & $-\log M_{\mathrm{He}}$ & $S$ (s) & Modes ($\ell,k$) \\
\hline
1. Circles (red) & 12\,100 & 0.795 & 4.5 & 2 & 0.047 & 109.33(1,1)\\
2. Triangles (blue) & {\bf 12\,400} & {\bf 0.79} & {\bf 4.5} & {\bf 2.5} &
0.001 & 109,28(1.1) \\ \hline
\end{tabular}
\caption{Absolute minima of the possible families of solution from the 
seismological analysis of G226-29.}
\label{tsol2}
\end{table}
 
\subsection{HL Tau76: rich man's seismology}

HL Tau76 is an example of a red edge pulsator, showing many periodicities. 
Because of the large number of modes, linear combinations and harmonics are
likely to appear. This can complicate the identification of the fundamental
modes. However, when daughter modes are detected, the parent modes have to be
excited as well, and the amplitude behavior of the daughter modes have to be 
similar to the parent modes. On the other hand, the dramatic observed
seasonal changes
changes in the pulsation spectra also make the search for parent modes more
difficult. Fortunately, the excited modes reappear at the same frequencies as in
the previous seasons. Considering that the transition zones between the layers 
are not sharp and that both convection zone depth
and temperature change during
the pulsation cycle (e.g. Montgomery 2005), a mode can appear at 600\,s and 
615\,s, depending on when the star is observed.

Another characteristic of red edge pulsators is the normally
high amplitude of the
observed modes. The ionization zone is getting deeper as the star cools down;
the energy transported by pulsations is larger. For these stars,
$T_{\mathrm{eff}}$ can vary of up to 500\,K (Robinson, Kepler, \& Nather 1982), 
almost half of the size of the ZZ Ceti instability strip ($\sim1\,200$\,K). 
Because of that, it is necessary to average out the stellar pulsation period 
when deriving an $T_{\mathrm{eff}}$ from spectral fluxes.

Table~\ref{tabhltau} shows the spectroscopic determinations for HL Tau76 and 
our estimative from the colors. We used the optical spectra determination as the
main guide, allowing our searches to be within three times the published 
uncertainties. This range includes also the color determinations.

\begin{table}
\begin{tabular}{||c|c|c|c|c||}\hline
Method & $T_{\mathrm{eff}}$ (K) & $\log g$ (in cgs) & $M$ ($M_{\odot}$)
& Spectroscopic reference \\ \hline \hline
LPT Optical spectra & 11\,450$\pm$200 & 7.89$\pm$0.05 & 0.55$\pm$0.02 &
Bergeron et al. 2004 \\ \hline
UBV colors & 10\,500$\pm$500 & 7.5$\pm$0.5 & 0.36$\pm$0.19 & This paper\\ \hline
\end{tabular}
\caption{Independent determinations for the atmospheric parameters for the star
HL Tau76.}
\label{tabhltau}
\end{table}

The list of HL Tau76 independent modes and their uncertainties calculated from
the amplitudes is shown in table~\ref{tab7}. The modes were published by
Dolez et al. (2006), and were detected in two WET campaigns. 

\begin{table}
\begin{tabular}{||c|c|c||}\hline
Periodicities (s) & Amplitude (mma) & Identification \\ \hline \hline
382.47$\pm$0.02 & 16.47 & $f_1$ \\
449.8$\pm$0.13 & 6.7 & $f_2$ \\
492.12$\pm$0.11 & 7.12 & $f_3$ \\
540.95$\pm$0.01 & 28.45 & $f_4$ \\
596.79$\pm$0.03 & 14.40 & $f_5$ \\
664.21$\pm$0.03 & 14.94 & $f_6$ \\
781.0$\pm$0.07 & 9.1 & $f_7$ \\
799.10$\pm$0.21 & 5.19 & $f_8$\\
933.64$\pm$1 & 2.40 & $f_9$ \\
976.38$\pm$0.14 & 6.46 & $f_{10}$\\
1064.97$\pm$0.05 & 11.30 & $f_{11}$ \\
1390.84$\pm$0.37 & 3.92 & $f_{12}$ \\ \hline
\end{tabular}
\caption{All detected modes for the star HL Tau76, in different years of
observations (Dolez et al. 2006).}
\label{tab7}
\end{table}

\begin{figure}
\resizebox{\hsize}{!}{\includegraphics[angle=0]{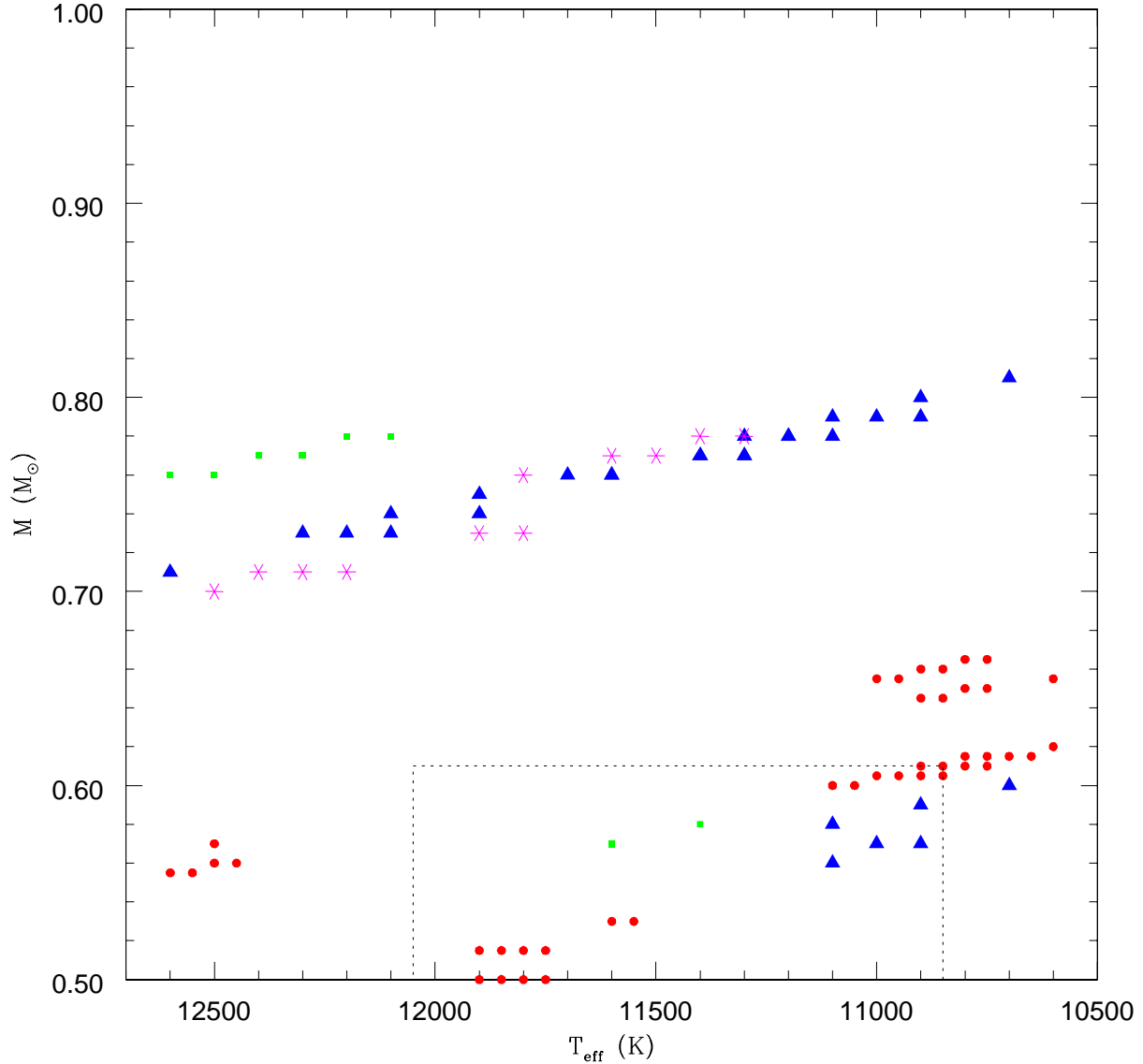}}
\caption{Results from the fit of the pulsation modes of the star HL Tau76 to
the models. The (red) circles are the solutions for 
$M_{\mathrm{He}}=10^{-2}\,M_*$, the (blue) triangles for 
$M_{\mathrm{He}}=10^{-2.5}\,M_*$, the (green) squares for 
$M_{\mathrm{He}}=10^{-3}\,M_*$, and the (magenta) asterisks for
$M_{\mathrm{He}}=10^{-3.5}\,M_*$. The dotted line box limits the region of the
independent temperature and mass determinations listed in table~\ref{tabhltau}.}
\label{sol4}
\end{figure}

We compared the observed modes with our model grid, assuming first 
that all modes
are $\ell=1$. The families of 
solutions are shown in figure~\ref{sol4}; in this case, the cut off is $S<5\,$s.
The minima of each family of solutions that is in agreement with the external
atmospheric determination are listed in table~\ref{tsol4}.

\begin{table}
\begin{tabular}{||l|c|c|c|c|c||}\hline
Symbol in the plot~\ref{sol2} & $T_{\mathrm{eff}}$ (K) & $M$ ($M_{\odot}$)
& $-\log M_{\mathrm{H}}$ & $-\log M_{\mathrm{He}}$ & $S$ (s) \\ \hline
1. Circles (red) & 11\,900 & 0.515 & 6 & 2 & 4.63 \\
2. Circles (red) & 10\,600 & 0.62 & 4 & 2 & 4.07 \\
3. Triangles (blue) & 10\,900 & 0.59 & 6.5 & 2.5 & 4.45 \\
4. Squares (green) & 11\,400 & 0.58 & 6.5 & 3 & 4.76 \\ \hline
\end{tabular}
\caption{Absolute minima for the various families of solutions from the
seismological analysis for the star HL Tau76, for the values closest to the
spectroscopic solutions.}
\label{tsol4}
\end{table}
                                                                                
The seismological study of the star HL Tau76 shows the difficulties of 
determining the internal characteristics of the star when its atmospheric 
parameters are not well established and when the star is close to the red edge, 
even though a large number of modes is excited. However, there is another piece
of information which has not been used for the hotter stars, but that is very
suitable for red edge pulsators: period spacing ($\Delta P$). For $k>5$ modes,
$\Delta P$ value will approach the asymptotic limit. This is not the case for
blue edge pulsators, where only low-$k$ modes are observed. From the expression:

\begin{equation}
\Delta P\propto\frac{1}{[\ell(\ell+1)]^{1/2}}
\end{equation}
$\Delta P$ is larger for $\ell=1$ modes than for $\ell=2$
($\Delta P_{\ell=1}/\Delta P_{\ell=2}=1.73$). In table~\ref{tabdeltap} we
list the differences in seconds between the modes.

\begin{table}
\begin{tabular}{||c|c||}\hline\hline
$\Delta P$ (s) & Modes \\ \hline
67 & $f_2-f_1$ \\
43 & $f_3-f_2$ \\
49 & $f_4-f_3$ \\
56 & $f_5-f_4$ \\
67 & $f_6-f_5$ \\
$2\times68$ & $f_7-f_6$ \\
18 & $f_8-f_7$ \\
$3\times46$ & $f_9-f_8$ \\
43 & $f_{10}-f_9$ \\
$2\times44$ & $f_{11}-f_{10}$ \\
325 & $f_{12}-f_{11}$ \\ \hline
\end{tabular}
\caption{Difference between consecutive modes for the star HL Tau76. The
$\Delta P\sim45$\,s is probably representative of differences between $\ell=2$
modes, while the $\Delta P\sim65$\,s for $\ell=1$.}
\label{tabdeltap}
\end{table}

\begin{figure}
\resizebox{\hsize}{!}{\includegraphics[angle=0]{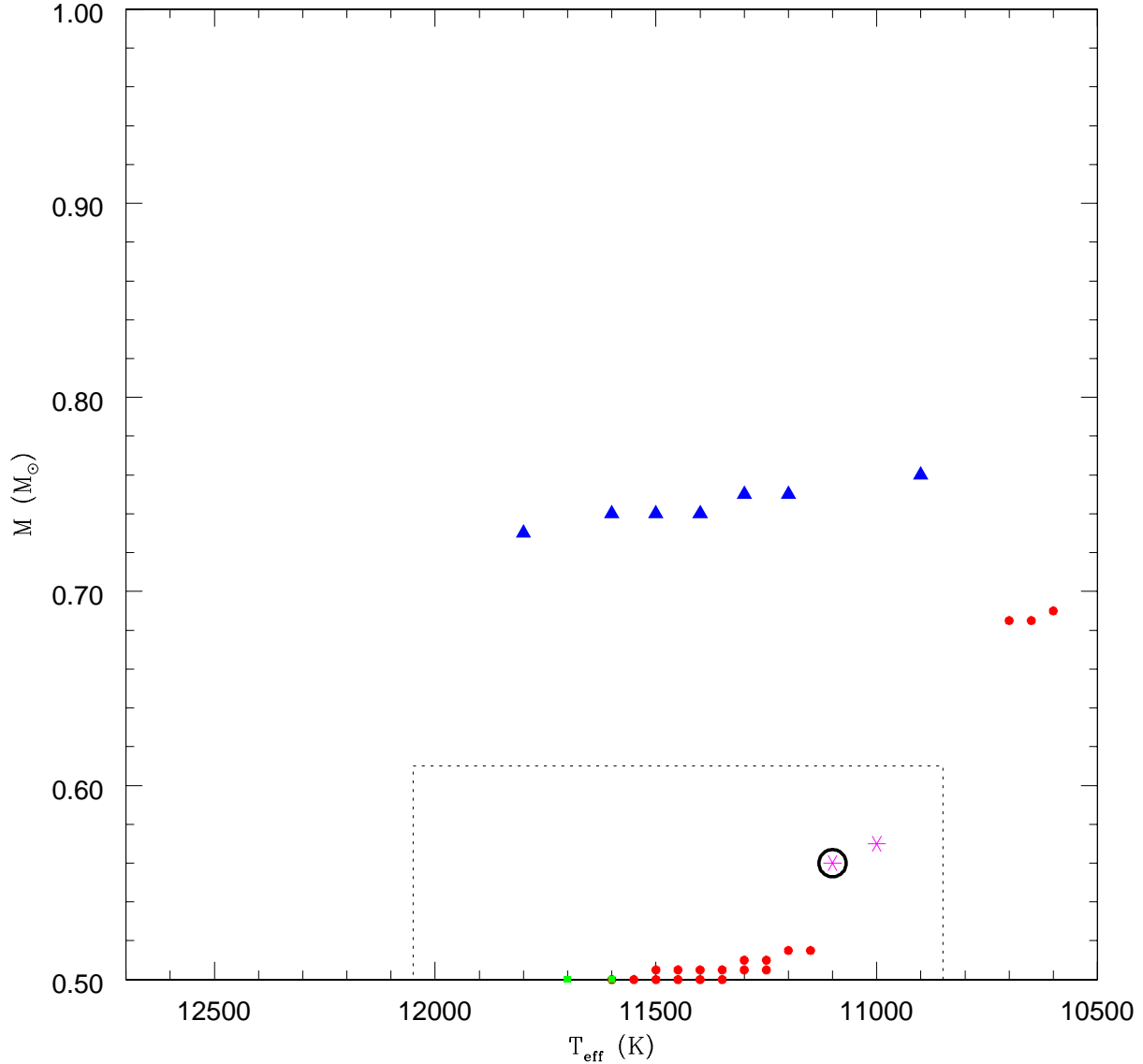}}
\caption{Results of the fit of the pulsation modes for the star HL Tau76
to the models. The (red) circles are the solutions for 
$M_{\mathrm{He}}=10^{-2}\,M_*$, the (blue) triangles for 
$M_{\mathrm{He}}=10^{-2.5}\,M_*$, the (green) squares for 
$M_{\mathrm{He}}=10^{-3}\,M_*$, and the (magenta) asterisks for 
$M_{\mathrm{He}}=10^{-3.5}\,M_*$. The higher amplitude modes and with period
spacing of $\sim$65\,s were fixed to as $\ell=1$ and the others could be fit
to $\ell=2$ modes. The dotted line box limits the region of the
independent temperature and mass determinations listed in table~\ref{tabhltau}
and the open circle shows the location of the minimum for this family of 
solutions.}
\label{sol4a}
\end{figure}

We found only two repeating values: $\sim45$ and $\sim65$\,s, which should be
representative of $\ell$=2 and 1 modes, respectively. Also, the $\Delta P$
between high amplitude modes is longer than for low amplitude modes, consistent
with the geometrical cancellation effects. The high amplitude modes should be
$\ell=1$, while the small amplitude ones, $\ell=2$. Therefore, we fixed the 
modes at 382.47\,s, 449.8\,s, 540.95\,s, 596.79\,s, 664.21\,s, 781.0\,s, and
1\,064.97\,s as $\ell=1$, and the other ones were allowed to fit either $\ell=2$
or 1. The families of solutions are plotted in figure~\ref{sol4a} and the
minima for each solution is in table~\ref{tsol4a}.

\begin{table}
\begin{tabular}{||l|c|c|c|c|c|c||}\hline
Symbol in plot~\ref{sol2} & $T_{\mathrm{eff}}$ (K) & $M$ ($M_{\odot}$)& 
$-\log M_{\mathrm{H}}$ & $-\log M_{\mathrm{He}}$ & $S$ (s) & Modes ($\ell,k$) \\
\hline\hline
1. Circles (red) & 10\,600 & 0.69 & 8 & 2 & 3.75 & \\\hline
2. Circles (red) & 11\,500 & 0.50 & 7.5 & 2 & 2.37 & \\ \hline
3. Triangles (blue) & 11\,500 & 0.74 & 5 & 2.5 & 3.39 & \\ \hline
4. Squares (green) & 11\,700 & 0.50 & 7.5 & 3 & 3.70 & \\  \hline
5. Asterisks (magenta) & {\bf 11\,100} & {\bf 0.56} & {\bf 8} & {\bf 3.5} &
3.20 & 395.9(1,3), 461.7(1,5)\\
&&&&&& 490,2(2,11), 539.8(1,6)\\
&&&&&& 596.4(1,7), 667.2(1,8)\\
&&&&&& 767.0(1,10), 789.0(2,20)\\
&&&&&& 924.3(2,24), 993.4(2,26)\\
&&&&&& 1054.9(1,15), 1397.5(1,21)\\\hline
\end{tabular}
\caption{Absolute minima of the various families of solutions from the 
seismological studies of HL Tau76, for the values close to the spectroscopic
solution. There are other families of solutions if the total mass is above
$0.70\,M_{\odot}$. The dotted line box represents the
spectroscopic determination in three sigma.}
\label{tsol4a}
\end{table}

The solutions 1 and 3 are for masses different than the spectroscopic value.
All the other solutions have a thin H layer. Solution 5 is the closest to 
previous determinations, with $T_{\mathrm{eff}}=11\,100$\,K, 
$M=0.56\,M_{\odot}$, $M_{\mathrm{H}}=10^{-8}\,M_*$, and 
$M_{\mathrm{He}}=10^{-3.5}\,M_*$.

\subsection{BPM37093: for millionaires}

BPM37093, a high mass white dwarf with 1.1$\pm0.05\,M_{\odot}$ (e.g. Bergeron
et al. 2004), was discovered to be a ZZ Ceti by Kanaan et al. (1992). Massive
white dwarfs have even higher densities and pressures than normal white dwarfs.
According to the evolutionary models, white dwarfs with $M>1\,M_{\odot}$ have a
significant crystallized portion already when they reach the ZZ Ceti instability
instability strip (Wood 1992, Winget et al. 1997, C\'orsico et al. 2005).

Like the stars at the red edge, BPM37093 shows irregularities in its pulsational
spectra. The nine independent modes detected by Kanaan et al. (2005), used in 
our seismological studies, are listed in table~\ref{tabbpm}.

\begin{table}
\caption[]{Modes for BPM37093}
\label{tabbpm}
\begin{tabular}{||c|c||}\hline
Periodicities (s) & Amplitudes (mma) \\ \hline \hline
511,7 & 0,68 \\
531,1 & 1,16 \\
548,8 & 0,98 \\
564,0 & 1,03 \\
582,0 & 1,03 \\
600,7 & 0,88 \\
613,5 & 1,13 \\
635,1 & 1,53 \\
660,8 & 0,48 \\ \hline\hline
\end{tabular}
\begin{list}{Table Notes.}
\item Detected modes in the star BPM37093, as listed in Kanaan et al. (2005).
\end{list}
\end{table}
                                                                                
Metcalfe, Montgomery, \& Kanaan (2004) labeled most of the
modes as $\ell=2$, due to the
small period spacing, $\langle \Delta P \rangle=17.6\pm1.1$\,s. They used fixed 
masses of 1.0, 1.03, and 1.1\,$M_{\odot}$, to search for the best temperatures
inside the instability strip, with a model-fitting method using a genetic 
algorithm. They explored only two possibilities of H and He layer masses: 
the range between $10^{-4}\,M_*$ and $10^{-8}\,M_*$ for H and 
$10^{-2}\,M_*$ and $10^{-4}\,M_*$ for He, and the
self-consistent results from massive stars evolutionary models, $10^{-5.8}\,M_*$
and $10^{-3.1}\,M_*$ (Althaus et al. 2003). They varied the crystallized portion
of a pure O nucleus. Their best fit indicates that 90\% of the stellar mass 
should be already crystallized.

There are only a few known high mass white dwarfs (Kepler et al. 2007), 
mostly because
they are rare, evolve fast, and have low luminosity. Therefore, the study of 
BPM37093 is very important to constrain the evolution of massive progenitors. Up
to date, it is the only $M>1\,M_\odot$ ZZ Ceti with reliable mass 
determinations.

Our initial model grid did not include masses higher than 1$\,M_{\odot}$, which 
is the case for BPM37093. As a first test, we searched for solutions inside the 
existing grid, but no solution was found even assuming that all modes are 
$\ell=2$. We expanded our model grid, including masses from 1.0\,$\,M_{\odot}$ 
to 1.1$\,M_{\odot}$, in 0.01$\,M_{\odot}$ steps. Because we were calculating
high mass models, we used a pure O core, as expected from self-consistent 
evolutionary calculations (Iben 1990, Iben et al. 1997).

Using the high mass grid, there is no solution if all modes were required to be
$\ell=1$. The period spacing and the low detected amplitudes are consistent with
$\ell=2$ modes. The minima of the families of the seismological solutions are
given in table~\ref{tabbpm2}.

\begin{table}
\begin{tabular}{||c|c|c|c|c|c||}\hline\hline
$T_{\mathrm{eff}}$ (K) & $M$ ($M_{\odot}$)& $-\log M_{\mathrm{H}}$
& $-\log M_{\mathrm{He}}$ & $S$ (s) & Modes ($\ell,k$) \\ \hline
11\,800 & 1.10 & 7 & 2 & 0.99 & 515.1(2,27), 531.7(2,28), 548.5(2,29), 
566.5(2,30)\\
&&&&& 582.6(1,17), 601.9(2,32), 613.1(1,18), 635.8(2,34) \\
&&&&& 653.9(2,35) \\ \hline
11\,700 & 1.08 & 6.5 & 2.5 & 1.02 & 514.8(2,27), 531.8(2,28), 549.0(2,29),
566.2(2,30) \\
&&&&& 584.0(2,31), 601.5(2,32), 614.2(1,18), 636.2(2,34) \\
&&&&& 653.6(2,35) \\ \hline
11\,600 & 1.07 & 6 & 3 & 1.03 & 514.8(2,27), 532.5(2,28), 549.5(2,29),
566.8(2,30) \\
&&&&& 583.2(1,17), 601.6(2,32), 612.9(1,18), 636.6(2,34) \\
&&&&& 654.0(2,35) \\ \hline
11\,800 & 1.10 & 6.5 & 3.5 & 1.15 & 514.5(2,27), 532.0(2,28), 549.7(2,29),
566.8(2,30)\\
&&&&& 584.3(2,31), 601.5(2,32), 614.6(1,18), 636.6(2,34)\\
&&&&& 654,2(2,35) \\ \hline
\end{tabular}
\caption{Absolute minima for the various families of solutions of the 
seismological analysis for the star BPM37093.}
\label{tabbpm2}
\end{table}

According to our seismological studies, the 613.5\,s mode, the third largest 
one, always fits better $\ell=1$, the 582.0\,s mode can be either $\ell=1$ or 2,
and all the other modes are better fitted by $\ell=2$. Our results are
in agreement with the analysis of Metcalfe, Montgomery, \& Kanaan (2004).

Even though it is not possible to determine the He layer mass, we can constrain
the total stellar masses as high ($M>1\,M_{\odot}$), temperatures between
11\,600 and 11\,800\,K (only 200\,K of external uncertainty), and H layer mass
between $10^{-6}\,M_*$ and $10^{-7}\,M_*$, consistent with seismological 
determinations from Althaus et al. (2003). Our seismological determinations for
the high mass red edge $T_{\mathrm{eff}}$ is also in theoretical and 
observational agreement with this to be hotter than the low mass red edge.

\section{Final discussions and conclusions}

In the present work we have built and explored an extensive model grid, which
calculates all possible modes that can be excited for a given internal structure
at certain temperatures. We have also developed an independent technique
of model fitting to compare the observed periods to the calculated periods. In
our approach, we used the external temperature and mass determinations to guide 
the seismological solutions, but we never limited the search to the uncertainty
range of the spectroscopic determinations. G117-B15A was the best star to test 
our seismological approach.

Our seismological study of a few ZZ Ceti stars is a clear evidence that 
seismology is really a powerful tool in the study of stellar evolution. Even for
the stars with few excited modes, it is possible to determine some 
characteristic of their interior. For G226-29, with only one detected mode it
was possible to restrict the mass to be above 0.7$\,M_{\odot}$. Combined with 
reliable values for temperature and mass, three internal parameters could be 
determined, because the modes do not show an asymptotic behavior, i.e., 
they are
more sensitive to
the structure of the star.

The study of G185-32 was the motivation to include the relative amplitudes to
give weights to the observed periods in the fits. The idea is that the high
amplitude modes should be present in the best models. Our conclusion is that
the observed amplitudes should be taken into account, even when one calculates
the modes from an adiabatic model.

As a red edge example, we analyzed HL Tau76. In this case, the period spacing
was a very useful tool, especially because of the asymptotic behavior of the
modes when they are larger. This analysis was consistent with the relative
amplitudes between the $\ell=1$ and 2 modes.

The success of the application of our approach to these five stars, including
also the well-studied BPM37093, encourages us to do seismology of all other
known ZZ Ceti stars. This is the topic of the second paper of this series, where
the results for the whole class will be presented and discussed.

\section*{acknowledgments}
We acknowledge support from the CNPq-Brazil. We also acknowledge Travis S.
Metcalfe for making available his scripts to calculate the model grid used in
this work and Agnes Bischoff-Kim for sharing her version of the model with
the Salaris profile.

\label{lastpage}
\end{document}